\documentclass[aps,prl,reprint,superscriptaddress,amsmath,amssymb,showpacs,showkeys]{revtex4-2}
\usepackage{graphicx}
\usepackage{dcolumn}
\usepackage{bm}
\usepackage{epstopdf}
\usepackage{color}
\usepackage[utf8]{inputenc}

\begin{document}
\title{Single-file dynamics of colloids in circular channels: time scales, scaling laws and their universality}
\author{Alejandro Villada-Balbuena}
\email{villadab@uni-duesseldorf.de}
\affiliation{Condensed Matter Physics Laboratory, Heinrich Heine University, Universit\"{a}tsstr. 1, 40225 D\"{u}sseldorf, Germany.}
\affiliation{Departamento de F\'isica, Cinvestav, Av. IPN 2508, Col. San Pedro Zacatenco, 07360 Gustavo A. Madero, Ciudad de M\'exico, Mexico.}
\author{Antonio Ortiz-Ambriz}
\affiliation{Departament de F\'{i}sica de la Mat\`{e}ria Condensada, Universitat de Barcelona, 08028 Spain}
\affiliation{Universitat de Barcelona Institute of Complex Systems (UBICS), Universitat de Barcelona, 08028 Spain}
\affiliation{Institut de Nanoci\`{e}ncia i Nanotecnologia, Universitat de Barcelona, 08028 Spain}
\author{Pavel Castro-Villarreal}
\affiliation{Facultad de Ciencias en F\'isica y Matem\'aticas, Universidad Aut\'onoma de Chiapas, Carretera Emiliano Zapata, Km. 8, Rancho San Francisco, 29050 Tuxtla Guti\'errez, Chiapas, Mexico.}
\author{Pietro Tierno}
\affiliation{Departament de F\'{i}sica de la Mat\`{e}ria Condensada, Universitat de Barcelona, 08028 Spain}
\affiliation{Universitat de Barcelona Institute of Complex Systems (UBICS), Universitat de Barcelona, 08028 Spain}
\affiliation{Institut de Nanoci\`{e}ncia i Nanotecnologia, Universitat de Barcelona, 08028 Spain}
\author{Ram\'on Casta\~{n}eda-Priego}
\affiliation{Divisi\'on de Ciencias e Ingenier\'ias, Campus Le\'on, Universidad de Guanajuato, Loma del Bosque 103, 37150 Le\'on, Guanajuato, Mexico.}
\author{Jos\'e Miguel M\'{e}ndez-Alcaraz}
\affiliation{Departamento de F\'isica, Cinvestav, Av. IPN 2508, Col. San Pedro Zacatenco, 07360 Gustavo A. Madero, Ciudad de M\'exico, Mexico.}
\date{\today}


\begin{abstract}
In colloidal systems, Brownian motion emerges from the massive separation of time and length scales associated to characteristic dynamics of the solute and solvent constituents. This separation of scales produces several temporal regimes in the colloidal dynamics when combined with the effects of the interaction between the particles, confinement conditions, and state variables, such as density and temperature. Some examples are the short- and long-time regimes in two- and three-dimensional open systems and the diffusive and sub-diffusive regimes observed in the single-file dynamics along a straight line. This work studies the way in which a confining geometry induces new time scales. We report on the dynamics of interacting colloidal particles moving along a circle by combining a heuristic theoretical analysis of the involved scales, Brownian Dynamics computer simulations, and video-microscopy experiments with paramagnetic colloids confined to lithographic circular channels subjected to an external magnetic field. The systems display four temporal regimes in this order: one-dimensional free diffusion, single-file sub-diffusion, free-cluster rotational diffusion, and the expected saturation due to the confinement. We also report analytical expressions for the mean-square angular displacement and crossover times obtained from scaling arguments, which accurately reproduce both experiments and simulations. Our generic approach can be used to predict the long-time dynamics of many other confined physical systems.
\end{abstract}
\maketitle


\section{Introduction}
Diffusion is one of the most common mechanisms used by nature to dissipate equilibrium density fluctuations, where the Brownian motion of colloidal particles represents a fascinating case \cite{Graham2018}. It exhibits a rich dynamical scenario in an extended time window due to the enormous separation of the characteristic time and length scales of solute and solvent constituents, combined with the effects of the interaction between the particles, the confinement conditions, and the state variables such as density and temperature \cite{gerhard1996,Dhont1996}. In particular, the way a confining geometry affects the colloidal diffusion has received much attention in the last decade \cite{CastroVillada2014,Omar2017}. However, from the experimental side, these effects are hard to study since they cover eight or more orders of magnitude, from milliseconds to several days \cite{CastroVillada2014,Zia2015,Taloni2017}. This feature makes it difficult to stabilize experimental set-ups over such a long time, and there are just few simulation results due to their high demand for computational power, which has held back theoretical developments as well.

The mean-square displacement (MSD or $W(t)$ in the following) of individual particles provides a good description of Brownian motion in systems where the possibility of particles finding a path among the other ones follows Gaussian diffusion \cite{gerhard1996,Dhont1996}. It also helps to describe the single-file diffusion (SFD) of particles moving along a straight line without mutual passage \cite{Kollmann2003}, a problem which is highly relevant in many scientific fields including biophysics and materials science \cite{Taloni2017}. Single-file diffusion of colloidal particles was first directly observed experimentally using paramagnetic or charged colloidal particles moving along a circle \cite{Wei2000, Lutz2004}, which found them to freely diffuse until reaching a subdiffusive regime, similar to the case of open and infinite straight lines \cite{Kollmann2003}. 

During the last few decades, an intensive research has been done in order to understand the universal fingerprint of the SFD behavior ($W(t) \sim\sqrt{t}$), where a plethora of theoretical explorations has been taken out for this purpose (see \cite{Taloni2017} for a review). However, since experiments and simulations have finite-size, new effects were reported when investigating confined systems characterized by a finite number of particles. Indeed, using a Bethe ansatz \cite{Shutz-1997}, an exact analytical expression for the tagged particle probability density function for hard-core interacting colloids in a finite box was found by Lizana et al. \cite{Lizana-2008} revealing the existence of three temporal regimes: normal diffusion, single-file diffusion and a saturation regime, which we define here as the geometric regime (GR). Particles interacting with a screened electrostatic potential confined in a narrow box were found to exhibit a good agreement with the expected scaling of SFD at intermediate times, whereas the GR was found at large times \cite{Delfau-2012}.  Additionally, the asymptotic behavior of the MSD was found to be $D_{N}t$, where $D_{N}= D_{0}/N$ for a finite system of $N$ colloids restricted to the line;  due to this characteristic collective behavior emerges, where the system behaves as single particle with an effective mass $N m$, with $D_{0}$ and $m$ being the free particle diffusion coefficient and the mass of each colloid \cite{Delfau-2011}, respectively.

All the above mechanisms have one common feature: they are caused by a geometrical and/or topological restriction. The tagged particle, as well as all interacting particles, experiences the effects of the geometry or topology imposed by the environment. In particular, the geometry can be manifested in various forms as the shape of the confinement (e.g., the parabolic confinement), the bounded domain of the narrow channel, the periodic structure of the substrate, the multilayer structure, the circular channel, among the many other forms one can find in nature. Geometry and topology, combined with the nature of the interaction between the particles, produce diverse types of effects \cite{Tarjus-2011}. For instance, topological defects, as kinks and anti-kinks, emerge when a colloidal monolayer is driven across commensurate and incommensurate substrate potentials \cite{Bohlein-2011}, as well as in highly dense systems of repulsive colloids in a narrow and periodic channel \cite{Siems-2015}. In some situations, curvature effects arise in the free diffusion processes over a curved manifolds \cite{Pavel2010}, where curvature becomes a fundamental physical quantity that acts just as an external field would do on the particles.
 
Recently, the original Ermak-McCammon algorithm for Brownian dynamics was extended to study Brownian motion of interacting particles confined on curved manifolds \cite{CastroVillada2014}. Particularly, the diffusion of a tagged particle without hydrodynamic interactions can be approached using the overdamped many-particle Langevin equation on an arbitrary plane curved file as follows \cite{CastroVillada2014},
\begin{eqnarray}
\zeta\frac{ds_{i}}{dt}=\sum_{i\neq j}F_{ij}^{T}+\eta_{i}, 
\label{MPLE}
\end{eqnarray}
where $\zeta$ is the friction coefficient, the sub-indexes labeled the particles, $s_{i}$ is the arc-length displacement, and $F_{ij}^{T}$ and $\eta_{i}$ are the interparticle force  and the stochastic force projected along the tangent direction at the $i$-th particle position, respectively. The stochastic force satisfies the fluctuation-dissipation theorem (see details in Ref. \cite{CastroVillada2014}). As a consequence of the tangent projection, all the dynamics occurs intrinsically along the curved file. In this case, the geometry encodes strong non-linear effects coupled to the interactions through the tangent projection. Indeed, paramagnetic colloids distributed along an ellipse were studied using Eq. (\ref{MPLE}) showing that curvature gradients induce inhomogeneities in the distribution of the particles along the file, and  providing evidence of metastable states through the behavior of the self-diffusion \cite{Omar2017}. Furthermore, the preliminary example of paramagnetic particles confined in a circular channel studied previously by some of us \cite{CastroVillada2014} provided evidence of two new temporal regimes beyond the sub-diffusive one not seen in experiments \cite{Wei2000,Lutz2004}, nor even in theoretical approaches for straight lines \cite{Kollmann2003}. Thus, it becomes evident that the colloid dynamics in a ring features a richer dynamical scenario that has not been studied in detail and will allow us to understand the role of the geometry on the dissipation of the equilibrium density fluctuations; a topic that has been overlooked so far.

Thus, by combining video-microscopy experiments performed with paramagnetic colloidal particles confined to lithographic circular microchannels subjected to an external magnetic field with the Ermack-McCammon algorithm for curved manifolds implemented for the paramagnetic colloids confined to a circle, we unravel the rich dynamical behavior of interacting colloidal systems that emerges due to the geometric confinement. Particularly, we focus on the angular distribution function and the mean-square angular displacement for several sets of the system parameters, namely, the number of particles, $N$, the radius of the circle, $R$, and the strength of the repulsive interaction between colloids, $\Gamma$. We find that the colloidal dynamics displays a universal behavior characterized by the following four temporal regimes: 1)free diffusion, 2)single-file diffusion, 3)free-cluster diffusion, and 4)geometrical diffusion. We provide accurate analytical expressions for the crossover times between these temporal regimes, which are obtained from scaling arguments, and derive accurate analytical representations for the mean-square angular displacement (MSAD) in terms of the crossover times.

In Sec. II, we present the experimental setups. In Sec. III, we describe the implemented Ermack-McCammon algorithm for curved manifolds. In Sec. IV, some experimental and simulation results for the structure and MSAD are shown. Secs. V and VI are dedicated to the time scales and scaling laws, respectively. In Sec. VII, we report precise analytical expressions for the MSAD. Finally, in Sec. VIII are some concluding remarks.

\section{Experimental setup}
We have performed video-microscopy experiments with paramagnetic colloidal particles confined to lithographic microgrooves in an external magnetic field following the procedure outlined in Ref. \cite{Ortiz2016a} with an improved protocol that will be discussed in the next paragraphs to ensure stability for longer periods of time. The experimental results obtained allow us to report the four temporal regimes and related time scales predicted with Brownian dynamics simulations.

Circular channels were first drawn on a Cromium (Cr) mask using direct write laser lithography ($\lambda = 405 \textrm{nm}$, $5-7\textrm{mm}^2\textrm{min}^{-1}$, DWL 66, Heidelberg Instruments Mikrotechnik GmbH). A thin coverglass ($\sim~120\mu\textrm{m}$) was coated with a $2.8\mu\textrm{m}$ layer of AZ-1512HS photoresist (Microchem, Newton, MA) by spinning it at 1000 r.p.m for 30s, and subsequently curing it at 95ºC for 3 minutes. To improve adhesion, before the photoresist, the glass surface was coated by a thin layer of TI Primer by spin coating for 20s at 4000 r.p.m and baking for 2 min at 120ºC. The photoresist was irradiated with UV light through the Cr mask for 3.4s at a power of 21mW cm$^{-2}$ (UV-NIL, SUSS Microtech), and then developed by submerging it for 45s in a solution of 1 part of AZ400K developer in 4 parts water, before washing with water. 

A colloidal suspension was prepared by mixing $1\mu l$ of superparamagnetic particles of diameter $\sigma=2.8\mu m$ (Dynabeads M-270, which consist of a polymer matrix embedded with iron-oxide nanoparticles) with $1\textrm{m}l$ of a 7$\mu$M solution of tetramethylammonium hydroxide (TMAH) in ultrapure water (Synergy UV-3, Millipore) at a pH of 7.2. TMAH has allowed us to avoid the absorption of $CO_{2}$ in water, which produces carbonic acid that glues the particles to the surfaces; this simple procedure enhanced the stability of the setup. A sample was prepared by sandwiching a droplet of solution between the structures and a top coverslip, separated and sealed with silicone vacuum grease (Dow Corning). The thickness of the sample is around 100$\mu$m. Before placing the top coverslip, the open sample is sonicated for 3 minutes to get rid of bubbles created by the superhydrophobicity induced by the concentric channels. The seal of the sample is essential to maintain the chemical stability of particles during the long experiments required to observe the different diffusion regimes. We have observed that a properly sealed sample is stable for about two weeks, before the dispersing medium becomes too acidic and electrostatic charges no longer are able to stabilize the particles over the substrate. During these long times, the sample is shielded from UV light to prevents further exposure of the photoresist. Particles have a density of around twice that of water ($\approx 2g \; ml^{-1}$) which is enough to keep them inside the channels. The particles are suspended by electrostatic interactions a few hundred nm above the surface.

\begin{figure}
\centering
\includegraphics[width=0.49\linewidth]{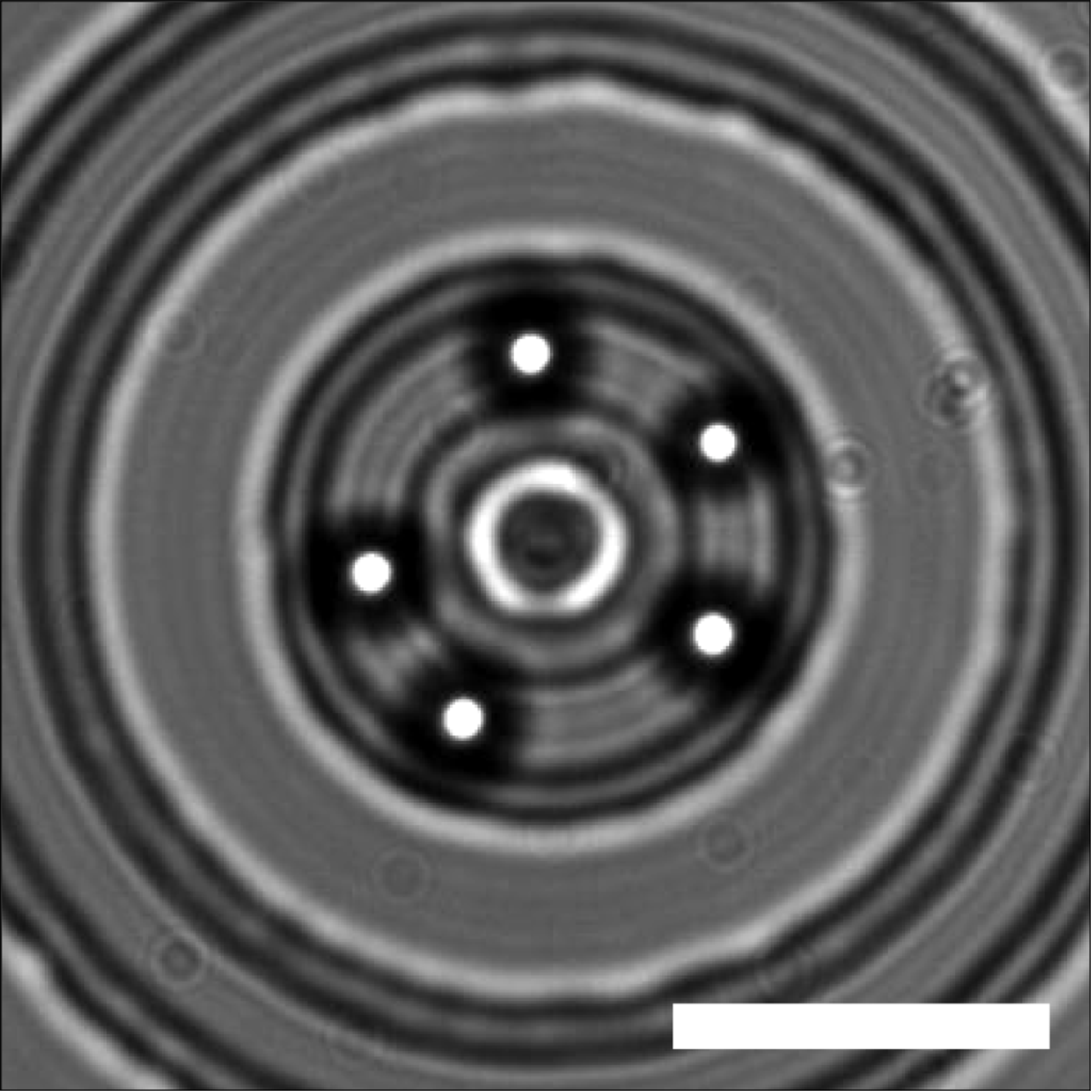}
\includegraphics[width=0.49\linewidth]{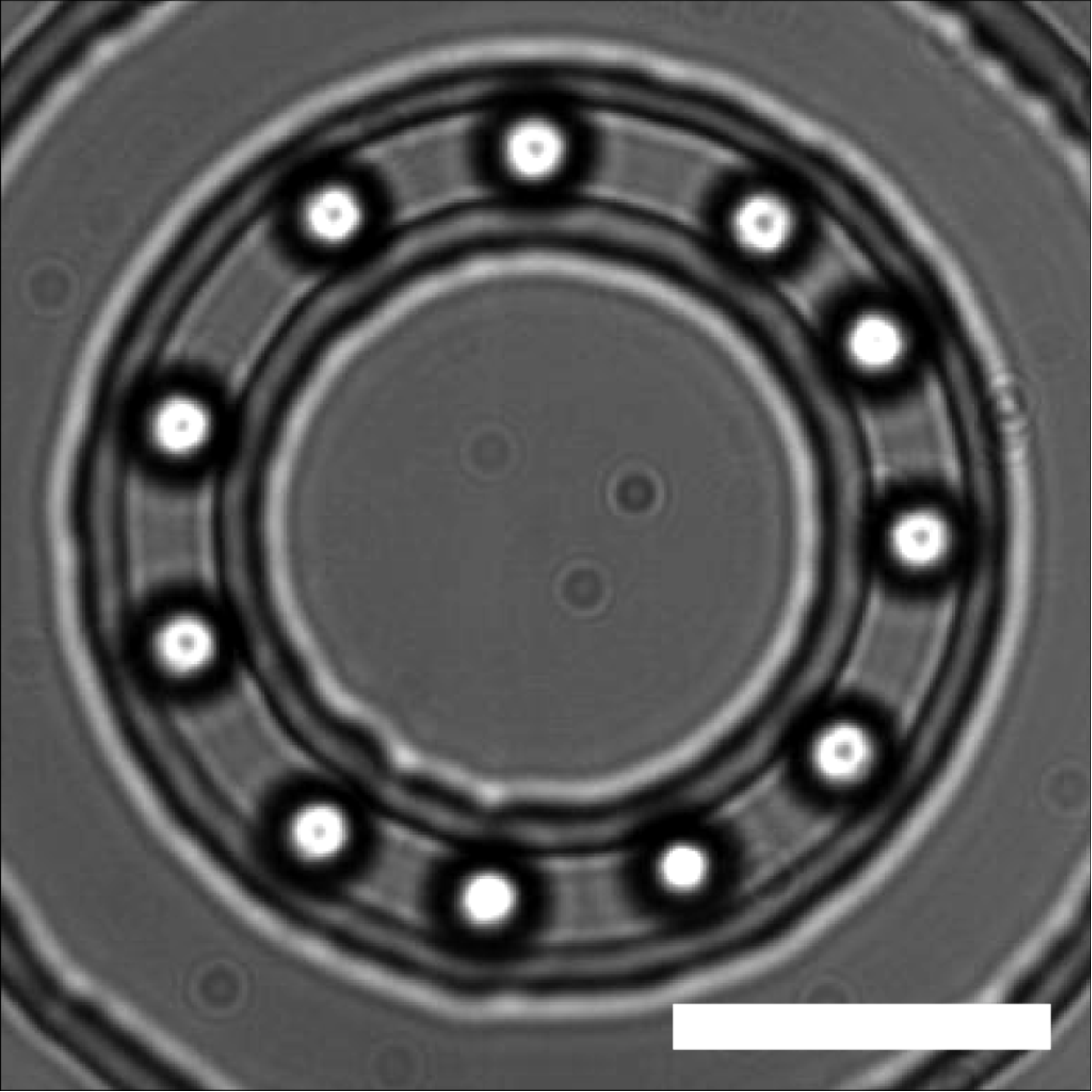}
\caption{Microscope image of colloidal particles confined by gravity in a microchannel. Left: channel radius $R=5\mu m$ and 5 particles. Right: channel radius $R=10\mu m$ and 11 particles. The widths of the channels are $2.5\mu m$ at the bottom and $5\mu m$ at the top, with a depth of $2.8\mu m$. The channel radius is measured from the center of the channel. Scale bar is 10$\mu m$ in both images.}
\label{Fig3}
\end{figure}

Once a sample is prepared, it is placed in a custom made optical microscope equipped with a 100x oil immersion objective (Nikon Plan Fluor, NA = 1.3, used with Thorlabs MOIL-30, $n=1.518$), a laser for optical trapping (ML5-CW-P-TKS-OTS, Manlight, 5W, operated at 3W) and a coil to apply a field along the axial direction (see Fig. \ref{Fig3}). The optical tweezer were used to move particles between the ring microchannels. The laser is deflected by an Acousto Optic Device (AOD AA Optoelectronics DTSXY-400-1064, driven by a radio frequency wave generator DDSPA2X-D431b-34 and a NI cDAQ card NI-9403) which allows fine control over the trap position and power. 
The time-sharing of laser light is especially important since magnetic particles absorb light at this wavelength, and they are prone to heating. We address this by keeping the power coming into the microscope objective between $1\textrm{m}W$ and $2\textrm{m}W$. To have the power approximately constant when moving the trap position, a telescope conjugates the plane of the AOD to the back focal plane of the objective. The system is operated by a custom made graphical interface programmed in LabVIEW, and available at \cite{Ortiz2020}. Observation was then done using a CMOS camera (Ximea MQ003MG-CM, 640$\times$480 pixels, pixel size is 7.4$\mu$m) working at a rate of 15fps for short timescales and 1fps for long timescales. From the videos, the particle positions are extracted using the trackpy implementation of the Crocker-Grier algorithm \cite{Crocker1996e}.

\section{Brownian dynamics simulations on the circle and superparamagnetic potential}

We use the Ermak-McCammon algorithm over curved manifolds developed in \cite{CastroVillada2014} in order to perform a systematical analysis of the interacting diffusing particles on the circle $S^{1}$.

According to Ref. \cite{CastroVillada2014}, the many-particle Langevin equation (\ref{MPLE}) on $S^{1}$ without hydrodynamic interactions can be rewritten as the following stochastic finite differences equation
\begin{equation}
s_i (t+\Delta t) = s_{i}(t) +  \beta D_0 {\bf F}_{i}\cdot{\bf T}_{i} \Delta t + \delta{\bf r}_i(\Delta t) \cdot {\bf T}_{i}, \label{eq1}    
\end{equation}
where $s_i(t)$ is the arc-length of the circle that can be computed using $s_i(t)=R \phi_i(t)$,  being $\phi_i(t)$  the angle measured counterclockwisely from the positive $x$-axis in the cartesian plane at time $t$. The total force, ${\bf F}_{i} = \sum_{i \not= j=1}^{N} {\bf F}_{ij}$, excerted on the $i$-th particle is the sum of  the forces,  ${\bf F}_{ij}$,  due to the interactions with all the particles $j\neq i$. In Eq. (\ref{eq1}), $\delta{\bf r}_i(\Delta t)$ is a Gaussian random displacement in the plane containing the circle with zero mean, $\langle\delta r_{i,\alpha}(\Delta t)\rangle = 0$, and covariance matrix $\langle\delta r_{i,\alpha}(\Delta t)\delta r_{j,\beta}(\Delta t)\rangle =2D_0 \delta_{ij} \delta_{\alpha \beta}\Delta t$, with $\delta r_{i,\alpha}(\Delta t)$ ($\alpha=1,2$) and $\left<\cdots \right>$ denoting, respectively, the Cartesian components of $\delta{\bf r}_i(\Delta t)$ and the ensemble averages. $\beta=(k_{B}T)^{-1}$ is the inverse of the thermal energy, $k_B T$, with $k_B$ and $T$ being the Boltzmann's constant and the absolute temperature, respectively. Additionally,  the unit tangent vector ${\bf T}_{i}=(-\sin \phi_i(t),\cos \phi_i(t))$ appearing in Eq. (\ref{eq1}) reflects the fact that the short-time dynamics occurs along the tangent line at the angle $\phi_{i}(t)$. According to those definitions, the arc-length displacement, $\Delta s_i(\Delta t) = s_i(t+\Delta t) - s_i(t)$ is related to the angular displacement $\Delta \phi_i(\Delta t) = \phi_i(t+\Delta t) - \phi_i(t)$ by $\Delta s_i(\Delta t)=R \Delta \phi_i(\Delta t)$, hence  the angular displacement is the sum between the deterministic part, $\beta D_0 {\bf F}_{i} \cdot {\bf T}_{i}$, and the random term,  $\delta{\bf r}_i(\Delta t)\cdot{\bf T}_{i}$.

In both experiments and simulations, we study a two-dimensional colloidal dispersion composed of $N$ paramagnetic particles with a magnetic moment $M$, whose pair potential is given by \cite{zahn1997}, 
\begin{equation}
\beta u(r_{ij}) = \frac{\mu_0}{4\pi k_B T} \frac{M^2}{r_{ij}^3}=\frac{\Gamma}{r_{ij}^{*3}} , \label{eq2}
\end{equation}
where $\mu_0$ is the vacuum magnetic permeability. For weak magnetic fields $M(B)=\chi_{eff}B$ holds, with $\chi_{eff}$ being the effective magnetic susceptibility of the colloids and $B$ the external magnetic field \cite{zahn1997}. The experiments were carried out at room temperature ($T=300 K$) with a magnetic susceptibility of $\chi_{eff}=(0.366 \pm 0.0002) \times 10^{-11} $Am${}^2 /$T. The right-hand side term in Eq. (\ref{eq2}) is the resulting dimensionless interaction potential, where $\Gamma$ is the total amplitude, and  $r_{ij}^{*}=r_{ij}/\sigma$ is the Euclidean distance between particles, expressed in terms of the particle dia\-me\-ter, $\sigma$.  As a result of the confinement and the finite size of the particles, the Euclidean distance, $r_{ij}=\sqrt{2}R\sqrt{1-\cos(\phi_i-\phi_j)}$, is bounded from above and below with maximum value $r_{ij}=2R$ when $\phi_i-\phi_j = \pm \pi$, and  minimum value $r_{ij}=\sigma$, when $\phi_i-\phi_j = \pm 2\sin^{-1}\left(\sigma/2R\right)$, respectively. Since the maximum angle between two particles is $\pi$, the domain of the angular distribution function, $g(\phi)$, is $[0,\pi]$. By using this feature, for instance, previous results \cite{viveros2008, Viveros2012} can be recovered  for both  dynamic and static properties.

Brownian dynamics simulations are carried out as follows. $N$ colloidal particles are set in random initial positions along the circle. Then, the colloidal system evolves according to Eq. (\ref{eq1}) from its arbitrary initial non-equilibrium state to the equilibrium one when the energy of the system reaches an average constant value, where the system is considered to be in equilibrium. The chosen time step for all the simulations is $\Delta t^{*} = \Delta t D_{0}/\sigma^{2} = 10^{-5}$. After reaching thermal equilibrium in $10^{6}$ time steps, we use at least $5 \times 10^{11}$ time steps to gather statistics, taking configurations every 100 time steps. In order to get good statistics at long times, we have parallelized the BD code to run 320 simulations of the same system, using a different time seed in each case, and averaging over all of them when simulations are done.

\section{Structure and Mean-Square Angular Displacement: Experiments \textit{vs} Computer Simulations}

In order to perform a one-to-one comparison between Brownian dynamics simulations and experiments, we choose three sets of experimental parameters, namely,  $\{N=5, R=5\mu m, \Gamma=1.07\}$, $\{N=5, R=5\mu m, \Gamma=4.26\}$, $\{N=5, R=5\mu m,\Gamma=9.59\}$ and $\{N=11, R=10\mu m, \Gamma=23.78\}$. In each case, $\Gamma$ is determined using the experimental parameters provided in the previous section. Also, different values for the magnitude of the magnetic field, $\{B[\text{mT}]=0.27, 0.54$, $0.8$, $1.27\}$, were used in each set.

\begin{figure}[htb!]
\centering
\includegraphics[width=\linewidth]{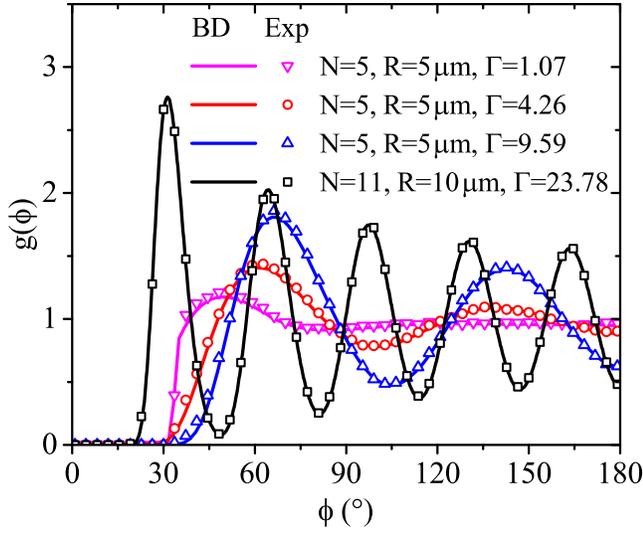}
\caption{Angular distribution function $g(\phi)$ for paramagnetic particles diffusing along a circle measured from video-microscopy experiments (open symbols) and calculated using BD simulations (solid lines) with the EM algorithm for curved manifolds (\ref{eq1}) for three sets of system parameters, as indicated in the graph.}
\label{Fig4}
\end{figure}

The angular distribution function, $g(\phi)$, i.e., the probability of finding a particle forming an angle $\phi$ with another particle, is explicitly shown in Fig. \ref{Fig4}. One can immediately notice that for all three sets of parameters there is an excellent agreement between experiments and simulations without using any free parameter, thus highlighting the accuracy of the modified Ermak-McCammon (EM) algorithm (\ref{eq1}) to reproduce the trajectory of interacting colloids on curved manifolds. Furthermore, as expected, we also observe that $g(\phi)$ becomes highly structured and long-range correlated when $\Gamma$ and $N$ increase. When $\Gamma = 1.07$, the paramagnetic repulsion is not enough to prevent the particles from colliding, as seen from the breaking of the pink line when reaching the correlation hole. We do consider volume exclusion in our simulations, but the effect is so slight that it does not have implications for the diffusion of the particles.

\begin{figure}[htb!]
\centering
\includegraphics[width=\linewidth]{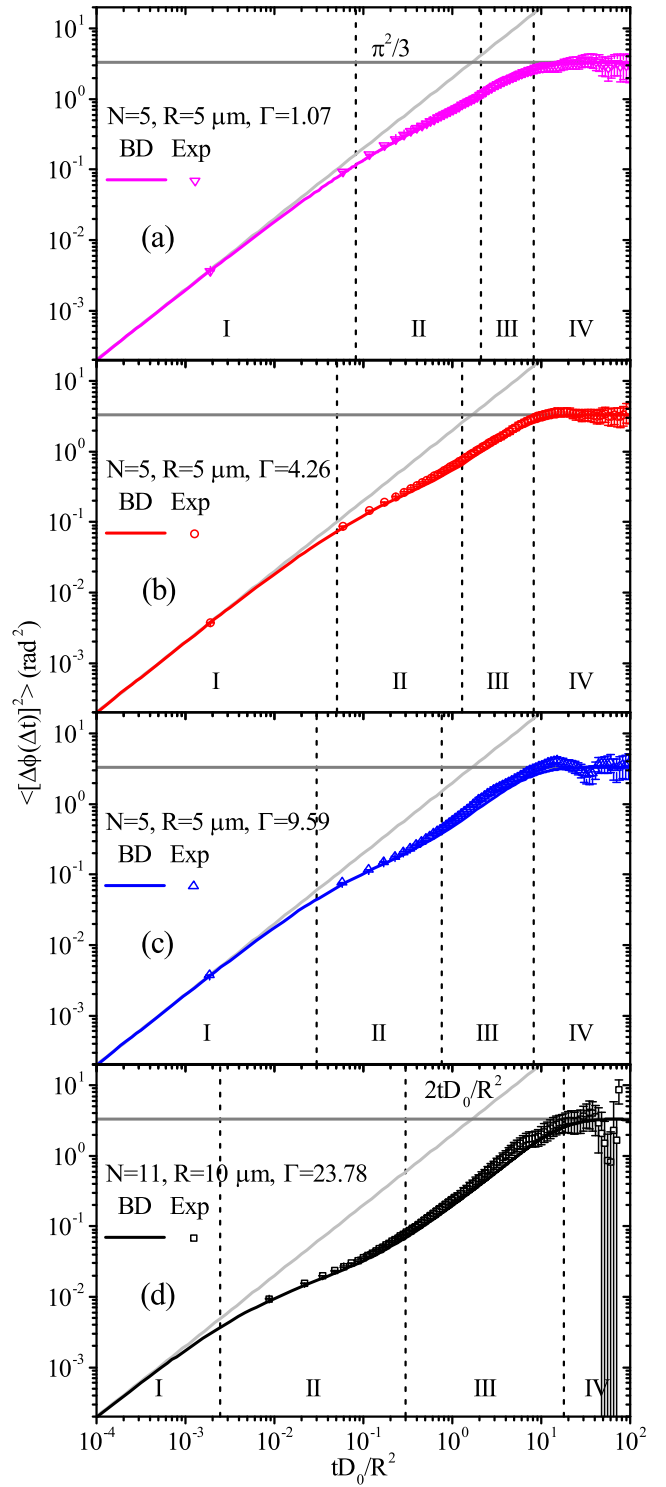}
\caption{Mean-square angular displacement $\langle [\Delta \phi (t)]^2 \rangle$ from experiments (open symbols) and BD simulations (solid lines). For all set of graphs, we identify the following four temporal regimes: the short-time diffusive regime (I), the sub-diffusive intermediate time regime (II), the second diffusive regime (III) and the geometrical saturation regime (IV). The straight gray lines are guides for the eye.}
\label{Fig5}
\end{figure}

In addition, in Fig. \ref{Fig5}, the Mean-Square Angular Displacement (MSAD) $\langle [\Delta \phi (t)]^2 \rangle$ is shown for the same set of system parameters as the one used for the calculations of the $g(\phi)$ displayed in Fig. \ref{Fig4}. Just as in a previous contribution \cite{CastroVillada2014}, we note that $\langle [\Delta \phi (t)]^2 \rangle$ exhibits systematically the following four temporal regimes: the short-time diffusive regime (I), the sub-diffusive intermediate time regime (II), the second diffusive regime (III) and, finally, the geometrical saturation regime (IV). From the figure, we also note an excellent agreement between experiments and BD simulations at short and first intermediate times, however, at the second intermediate time regime one can notice small differences that slightly increase with $\Gamma$. Overall, the agreement is outstanding given that there are no free parameters in the calculations. Afterwards, at the geometrical regime, a larger difference is evident for $N=11$ as a consequence of a lack of statistics in the experiment, mainly because of the difficulty to keep the measurement stable for very long times.

\section{Time Scales from the Mean-Square Angular Displacement}

Given the excellent agreement between experiments and Brownian dynamics simulation results discussed previously, we now proceed to explore additional sets of the system parameters fixing in each case two of the following three physical variables ${N, R^{*}=R/\sigma, \Gamma}$, and varying the remaining one,
\small\begin{enumerate}
\item \{$R^{*}=10.5$; $\Gamma=125$; $N=1,10,15,20,25,32$\}
\item \{$N=26$; $\Gamma=125$, $R^{*}=10.5, 13, 15.5, 18, 20.5, 30.5$\}
\item \{$N=10$; $R^{*}=10.5$; $\Gamma=10, 10^2, 10^3, 10^4$\}
\end{enumerate}

\begin{figure}[htb!]
\includegraphics[width=\linewidth]{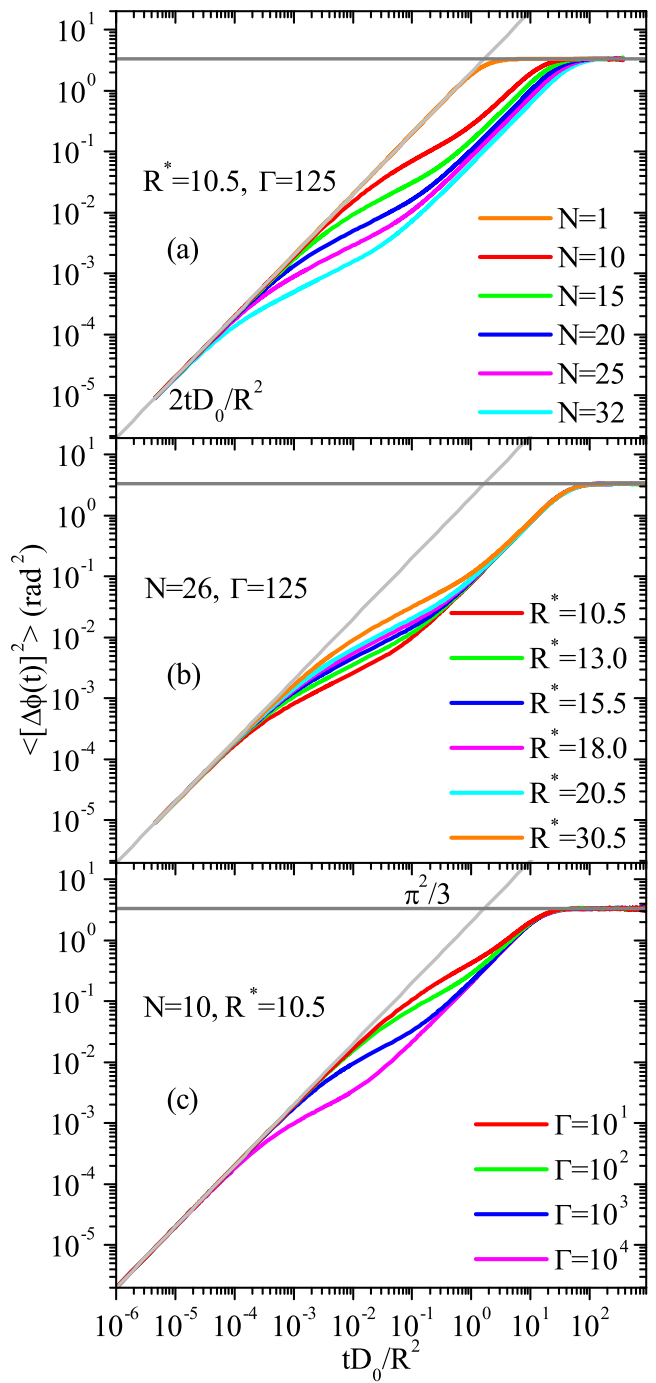}
\caption{Mean-square angular displacement $\langle [\Delta \phi (t)]^2 \rangle$ obtained with the EM algorithm (\ref{eq2}) for paramagnetic colloids diffusing along a circle. (a) For different number of particles, (b) varying the radius of the circle and (c) changing the strength of the inter-particle repulsion, as indicated. The thin straight lines are guides for the eyes, showing the short-time, $2D_{0} t/R^2$, and the geometrical limit, $\pi^2/3$, values.}
\label{fig1}
\end{figure}

Results for the mean-square angular displacement are shown in Fig. \ref{fig1}. It turns out that the four time-regimes depicted in Fig. \ref{Fig5} are also observed in the dynamical behavior of the MSAD of Fig. \ref{fig1}. In order to better understand the dynamical behavior at each regime, we will now classify all of them in terms of the characteristic transition times, $\tau_i$, that define their beginning and the end: short time regime $t\leq \tau_{d}$, first intermediate time regime $\tau_{d}\leq t\leq \tau_{c}$, second intermediate time regime $\tau_{c}\leq t\leq\tau_{G}$, and geometrical regime $t > \tau_{G}$,  where $\tau_{d},\tau_{c}$, and $\tau_{G}$ depend clearly on the physical parameters $N, R^{*}$ and $\Gamma$. One of the main goals of this contribution is to express the dependence of these transition times on the aforementioned physical quantities. We now explain the behavior of the MSAD in each of these regimes. 

According to the results shown in Fig. \ref{fig1}, at the short time regime, the MSAD  displays a $1D$ free-particle-like behavior,
\begin{equation}
\langle [\Delta \phi (t)]^2 \rangle = 2 \frac{D_{0}}{R^2}  t.
\label{eq4}
\end{equation}
This behavior states that, at short times, the deterministic term, $\beta {D}_{0} {\bf F}_{i}\cdot{\bf T}_{i} \Delta t$, is negligible in comparison to the random term,  $\delta{\bf r}_{i}(\Delta t) \cdot {\bf T}_{i}$ in the the Ermak-McCammon algorithm (\ref{eq1}). This implies that, at short times, fluctuations play an essential role in contrast to the interactions between the particles. Also, a $1D$ Brownian motion is found, since the circle is a one dimensional manifold, which can be considered locally as a straight line. 

In the first intermediate regime, the interactions between the particles take place and this results in a deviation of the MSAD from the free-particle behavior to the dynamical transition known as single-file diffusion,
\begin{equation}
\langle \left[ \Delta \phi  (t) \right]^2 \rangle = F \sqrt{t}/R^2, \label{Feq}
\end{equation}
where $F$ is the single file (SF) mobility, and the dependence $\sqrt{t}$ is inferred from the slope of the curve in Fig. \ref{fig1}. SFD occurs because of the excluded mutual passage between the particles \cite{Taloni2017}. Consequently, this phenomenon appears also for any type of repulsive interaction, even for the simplest hard core ones, as long as the particles are kept confined in the single line \cite{Taloni2017}. Recall that the transition to the SFD occurs sooner as the particle density increases, which is achieved when either $N$ increases or $R$ decreases. According to the prediction reported in Ref. \cite{Kollmann2003}, the coefficient $F$ depends on the relative compressibility. In our case, we will show below how $F$ depends on the strength of the interaction potential.

In the third regime, the curves of the MSAD are basically parallel to the curve of the free-particle behavior, which indicates that the  MSAD is again proportional to time $t$. In addition, it turns out that the proportionality coefficient is $2D_{0}/N$. Thus, the dynamical behavior at the second intermediate time regime can be expressed as follows,
\begin{equation}
\langle \left[ \Delta \phi  (t) \right]^2 \rangle = 2 \frac{D_{0}}{N R^2} t.
\label{eq8}
\end{equation}
Noticeably, this result highlights the fact that the whole bunch of particles confined in the circle behaves as if the system would be just one particle in a fluid with an effective drag coefficient $\zeta_{N} = N\zeta$, where $\zeta=k_{B}T/D_{0}$ is the drag coefficient between any of the colloids and the fluid. As a result, at this time regime, a collective behavior emerges where the whole system acts as one ring shaped particle with an effective diffusion coefficient $D_{N}=D_0/N$ performing a rotational Brownian motion. In other words, the particles organize in such a way that they randomly move together as a cluster, and again the collective dynamics is independent from the interaction potential. This phenomenon is a consequence of the finite-size of the system imposed by the topology of the circle and it does not have counterpart for the case of open systems. A similar situation appears in the case of the finite-size colloidal system confined in a line \cite{Delfau-2011}. We expect this kind of transition to be universally valid for any repulsive interaction potential as long as the particles are constrained along a circle \cite{villada2018}. In other words, we find a new transition from the single file diffusion to the cluster file diffusion. 

Finally, in the fourth time regime, the system reaches the so called geometric regime, where particles have explored the positions of the circle many times in such a way that any colloid has the same probability to be found at any position on the circle, independently of the interaction between the particles. This result was previously reported in Ref. \cite{CastroVillada2014} and has an analytical value,
\begin{equation}
\langle \left[ \Delta \phi  (t) \right]^2 \rangle = \frac{\pi^{2}}{3}.
 \label{eq9}
\end{equation}

Thus, from the previous discussion and the dynamical behavior reported in Fig. \ref{fig1}, one can appreciate that the only parameters undetermined are the single-file mobility $F$, and the transition times $\tau_{d}, \tau_{c}$ and $\tau_{G}$ that we determine in the following.

\section{Crossover times: scaling laws}
\begin{figure}[htb!]
\includegraphics[width=0.98\linewidth]{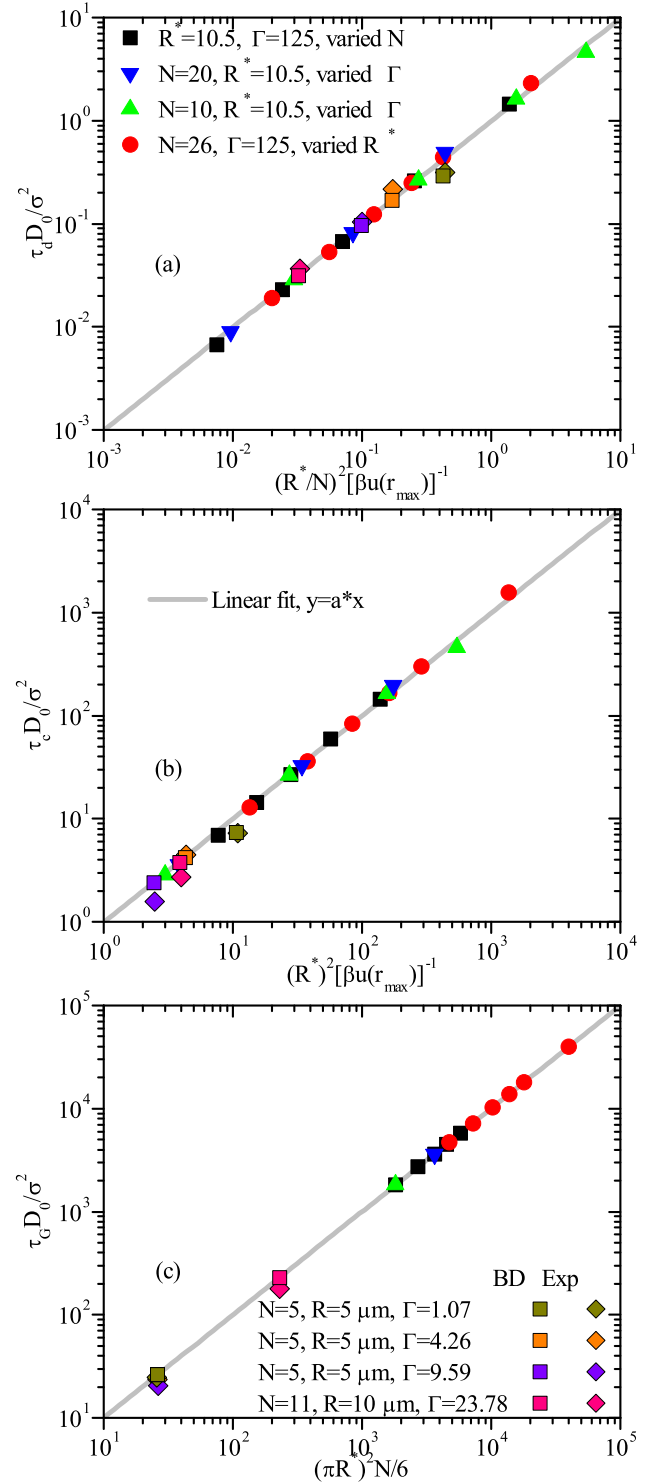}
\caption{Transition times $\tau_{d}, \tau_{c}$ and $\tau_{G}$ for the crossover between the different regimes for paramagnetic colloids diffusing along a circle of radius $R$. In (a) and (b) the straight line is the linear fit with slope $a \approx 0.99 \pm 1.74\times10^{-2}$.  The line in (c) is the best linear fit with slope  $a \approx 1.0 \pm 2.78\times10^{-14}$. Both experimental and simulation results displayed in Figs. \ref{Fig5} and \ref{fig1} are included.}
\label{fig2}
\end{figure}

We now provide a deduction of the SF mobility and the transition times $\tau_{d}, \tau_{c}$ and $\tau_{G}$. First, let us notice that from the Ermak-McCammon algorithm (\ref{eq1}) we are able to show that,
\begin{eqnarray}
\langle \left[ \Delta s_{i} (\tau) \right]^2 \rangle  = \langle \left[ \beta D_{0} \tau {\bf F}_{i} \cdot  {\bf T}_{i} \right]^2 \rangle + 2 D_{0} \tau, \label{eq22}
\end{eqnarray}
by squaring (\ref{eq1}) and taking its average. The crossed product does not appear in Eq. (\ref{eq22}) since the deterministic term $\beta D_0 {\bf F}_{i}\cdot{\bf T}_{i} \tau$ and the tangent vector ${\bf T}_{i}$ in the random term both depend on the previous time $t$ respect to $t+\tau$. Clearly, when interactions are absent, the last equation reproduces the behavior of the MSAD at the short-time regime (\ref{eq4}). In addition, when inter-particle interactions are considered, the first term of (\ref{eq22}) becomes important. Indeed, this term has the same order of magnitude of the second one when the system reaches the single file diffusion regime. 

In order to determine the first transition time, first, let us carry out a dimensional analysis, where we denote by $[\mathcal{Q}]$ the dimension of the quantity $\mathcal{Q}$. Hence,  the deterministic term $\beta D_0 {\bf F}_{i}\cdot{\bf T}_{i} \tau$ in the Ermak-McCammon algorithm (\ref{eq1}) has dimension of length. Thus, let us denote by $\ell$ a characteristic length to be inferred from the simulation. Then, $\ell=\beta D_{0}\left[{\bf F}_{i}\cdot{\bf T}_{i}\right]\tau$, where $\tau$ is a characteristic time associated to $\ell$.  The tangent vector ${\bf T}_{i}$ does not have any dimensions, thus $\left[{\bf F}_{i}\cdot{\bf T}_{i}\right]=\mathcal{U}/\ell$, where $\mathcal{U}$ has units of energy. Combining all these terms, one gets the following expression,
\begin{eqnarray}
\frac{D_{0}\tau}{\sigma^2}=\mathcal{C}\left(\frac{\ell}{\sigma}\right)^2\left(\beta\mathcal{U}\right)^{-1}.
\label{TransTime}
\end{eqnarray}
In this equation, there are three quantities to be determined, namely, the characteristic length $\ell$, the value of $\mathcal{U}$ and the dimensionless constant $\mathcal{C}$. As it was explained earlier, the interaction term becomes relevant at the intermediate time regime, therefore, the first and second transition times should have the form shown in Eq. (\ref{TransTime}). The transition to single file behavior occurs when a particle approaches another one, such that the interaction becomes important and prevents mutual passage, thus this should occur at the angle $\phi_{\rm max} $, which corresponds to the first peak of  $g(\phi)$ that gives the maximum probability that two particles encounter each other. Additionally, it is known that $g(\phi)$ carries information of the interaction potential. Therefore, one could infer that for the transition times (\ref{TransTime}) $\mathcal{U}=u(r_{\rm max})$, where $r_{\rm max}=\sqrt{2}R\left(1-\cos\phi_{\rm max}\right)^{1/2}$. Now, in order to determine the value of $\ell$, we have performed perturbation theory on the Ermak-McCammon algorithm (\ref{eq1}) and found that for the first transition time, it has the form $D_{0}\tau_{d}/\sigma^2= \mathcal{C}_{\rm th} \left(R^{*}/N\right)^2\left(\beta u(r_{\rm max})\right)^{-1}$, where the theoretical value  for the dimensionless constant is $\mathcal{C}_{\rm th}\approx 1.3$ (see Appendix \ref{app}). From this analysis,  it turns out that $\ell=R/N$ and according to the simulation result displayed in Fig. (\ref{fig2})  $\mathcal{C}\approx 1$. Thus, one has the following mathematical expression for $\tau_{d}$,
\begin{equation}
\frac{D_{0}\tau_{d}}{\sigma^2}=\left(\frac{R^{*}}{N}\right)^2\left(\beta u(r_{\rm max})\right)^{-1}.
\label{eq10}
\end{equation}
This expression has an excellent agreement with both experimental and simulation results as it is shown on the top plot of Fig. \ref{fig2}.

\begin{figure}[htb!]
\centering
\includegraphics[width=\linewidth]{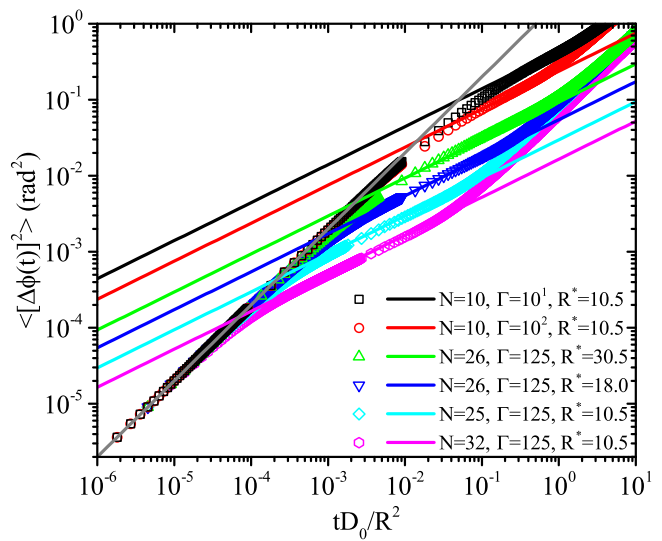}
\caption{Mean-square angular displacement for six paramagnetic colloidal systems obtained with the EM algorithm (\ref{eq1}). Bold lines represent Eq. (\ref{Feq}) with the mobility factor given by Eq. (\ref{eq11}) evaluated using the physical parameters as indicated in the graph.}
\label{figF}
\end{figure}

The deduction of the SF mobility $F$ can be carried out using $\tau_{d}$, Eq. (\ref{eq4}) and Eq. (\ref{Feq}). This can be obtained by equating Eq. (\ref{eq4}) for the free-particle diffusion and Eq. (\ref{Feq}) for the single-file diffusion at the first transition time $\tau_{d}$, namely, $2D_{0}\tau_{d}=F\sqrt{\tau_{d}}$, that is
\begin{equation}
F=\left( \frac{2 R}{N} \right) \sqrt{\frac{D_{0}}{\beta u(r_{max})}}.
\label{eq11}
\end{equation}
This expression is straightforward and reproduces correctly both the experiments and simulations (see Figs. \ref{fig2}a and \ref{figF}). In fact, from Fig. \ref{figF}, one can see the remarkable good agreement between expressions  (\ref{Feq}) and (\ref{eq11}) when compared with the simulation results. As pointed out above, Kollmann \cite{Kollmann2003} derived an expression for $F$ in terms of the compressibility or the evaluation of the structure factor at $q\rightarrow 0$. However, in practice, that route needs the simulation of larger system sizes and, consequently, demands a high computational cost. Thus, making Eq. (\ref{eq11}) more feasible to explain the transport of particles when the SFD condition is reached. 

Now, using Eq. (\ref{Feq}) with the value for $F$ given by Eq. (\ref{eq11}), one can proceed to deduce the second transition time, $\tau_{c}$, by equating (\ref{Feq}) and (\ref{eq8}). Hence, one now obtains that,
\begin{eqnarray}
\frac{D_{0}\tau_{c}}{\sigma^2}=R^{*2}\left(\beta u(r_{\rm max})\right)^{-1}.
\label{eq12}
\end{eqnarray}

As it can also be seen in Fig. \ref{fig2}b, the previous expression for $\tau_{c}$ shows an excellent agreement between experiments and simulation results. The last transition time, $\tau_{G}$, can be easily obtained by equating Eqns. (\ref{eq8}) and (\ref{eq9}), that is,
\begin{equation}
\frac{D_{0}\tau_{G}}{\sigma^2} = \frac{N}{6} (\pi R^{*})^2 .
\label{tauG}
\end{equation}
Fig. \ref{fig2}c shows clearly that the previous expression describes remarkable well both experimental data and simulation results. We obtained the experimental and simulation points in Figure \ref{fig2} as follows. Their vertical values from the crosses between the respective linear fits (on a log-log scale) of the different temporal regimes in Figures \ref{Fig5} and \ref{fig1}. Their horizontal values using the position of the first maxima of the angular distribution functions shown in Figure \ref{Fig4} in the expression for the interaction potential.

\section{Asymptotic behavior}

While one of the novelties of this work is the expression for the single file mobility (\ref{eq11}), another immediate result is the availability of writing the subdiffusive regime as a function of the first transition time,
\begin{eqnarray}
\langle \left[ \Delta \phi (t) \right]^2 \rangle =\frac{F}{R^2} \sqrt{t} = 2 \frac{D_0}{R^2} \sqrt{\tau_d t}.
\label{eq13}
\end{eqnarray}
By doing this, we notice that Eq. (\ref{eq13}) is similar to the diffusive behaviors described in Eqns. (\ref{eq4}) and (\ref{eq8}), which give us the possibility to construct a relationship that connects all the different time regimes in terms of the crossover times: the first diffusive regime ($\langle \left[ \Delta \phi (t) \right]^2 \rangle \sim t$), the subdiffusive regime ($\langle \left[ \Delta \phi (t) \right]^2 \rangle \sim \sqrt{t}$), the second diffusive regime ($\langle \left[ \Delta \phi (t) \right]^2 \rangle \sim t/N$) and the geometrical regime ($\langle \left[ \Delta \phi (t) \right]^2 \rangle = \pi^2 /3$). To this end, we first propose an equation that describes both the first diffusive and the subdiffusive regimes given by Eqns. (\ref{eq4}) and (\ref{eq13}), respectively. We then use an exponential decay associated to the crossover time $\tau_d$,
\begin{eqnarray}
\langle \left[ \Delta \phi (t) \right]^2 \rangle =2 \frac{D_0}{R^2} t \left[ 1-\exp\left( -\frac{\tau_d}{t} \right) \right]^{1/2}.
\label{eq14}
\end{eqnarray}
This equation indeed displays the diffusive behaviour at short times, while at longer times it will indefinitely describe the subdiffusive one, recovering both time regimes.

Analogously, for the second transition, we now propose another exponential decay, but in this case the transition is from $\sqrt{t}$ to $t$ linked to the crossover time $\tau_{c}$,
\begin{eqnarray}
\langle \left[ \Delta \phi (t) \right]^2 \rangle =2 \frac{D_0}{R^2} t \left[ \frac{ 1-\exp\left( -\frac{\tau_d}{t} \right) }{ 1-\exp\left( -\frac{\tau_c}{t} \right) } \right]^{1/2}.
\label{eq15}
\end{eqnarray}
If the angle was not bound, this equation would  describe correctly the mean-square angular displacement (data not shown). However, we have used the bounds convention, i.e., $\phi \in \left[ 0,\pi \right]$. With this condition, the last transition time occurs when the mean-square angular displacement reaches a plateau, changing the linear dependence with time to a constant value. This dynamical feature can be included in Eq. (\ref{eq15}) by considering a term that recovers the geometrical time regime as follows,
\begin{eqnarray}
\langle \left[ \Delta \phi (t) \right]^2 \rangle &=&2 \frac{D_0}{R^2} t \left[ \frac{1-\exp\left( -\frac{\tau_d}{t} \right)}{1-\exp\left( -\frac{\tau_c}{t} \right)} \right]^{1/2} \times\nonumber\\
& & \left[ 1-\exp\left( -\frac{\tau_G^2}{t^2} \right) \right]^{1/2}.
\label{eq16}
\end{eqnarray}

\begin{figure}[htb!]
\centering
\includegraphics[width=\linewidth]{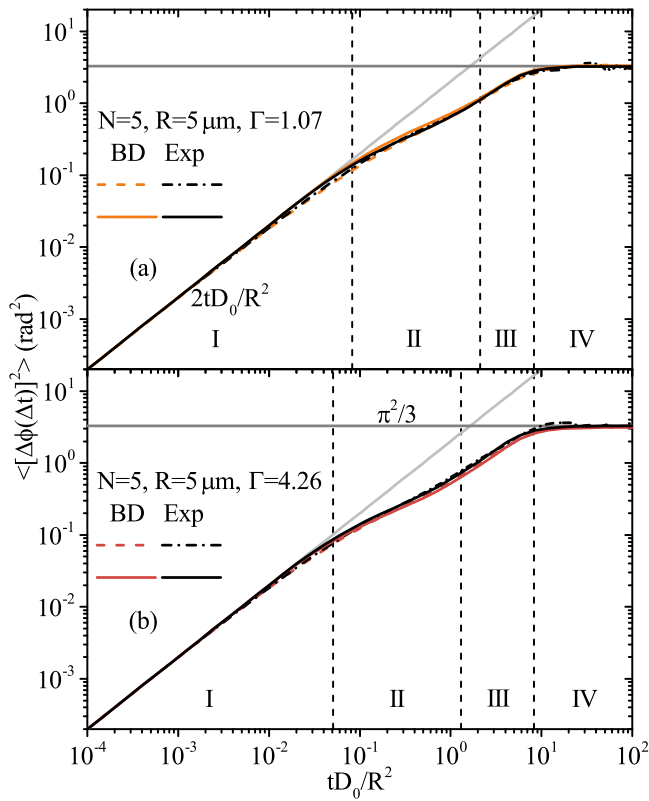}
\caption{Comparison of the mean-square angular displacement for the systems displayed in Fig. \ref{Fig5} (a) and (b) (dotted lines, red for BD and black for experiments) and Eq. (\ref{eq16}) (continuous lines with the same color code) using the crossover times reported in Fig. \ref{fig2}.}
\label{AnSol}
\end{figure}

In order to test the accuracy of this equation, the crossover times reported in Fig. \ref{fig2} were used in such expression and the resulting MSAD was compared with its corresponding counterpart. In all cases, the agreement is rather good. For illustrative purposes, in Fig. \ref{AnSol} we only show a comparison between the results reported in Figs. \ref{Fig5} (a) and (b) and Eq. (\ref{eq16}).

\section{Concluding remarks}

In this contribution, we have studied the dynamical behavior of paramagnetic particles confined to move along a circle. Although this system was already studied in Refs. \cite{Wei2000, Lutz2004}, we demonstrated here new features in the particle dynamics when the circle geometry is taken into account; those features are also corroborated with experiments performed at very long times that clearly reach the predicted geometrical time regime. This study highlights that small and closed systems exhibits a richer dynamical behavior than the standard scenario of single-file diffusion along a straight and infinite line, where the mean-square displacement behaves, at long times, as $\sim \sqrt{t}$.

Furthermore, we provide evidence on a firm ground of the existence of four temporal regimes of the stochastic dynamics of a tagged particle confined along a circle. Although these regimes have been previously studied only numerically \cite{CastroVillada2014}, here we provide a complete picture of the phenomenon, which includes experiments, simulations and a predictive analytical theory. These regimes are described in terms of the mean-square angular displacement, where a good consistency is found between the experimental measurement and the corresponding theoretical prediction. The first and second temporal regimes corresponds to the usual transition from free particle diffusion to the single file behavior, respectively, since the particles cannot pass through each other. However, after the single-file regime, a new transition to a diffusion type regime emerged with a reduce diffusion coefficient given by $D_{0}/N$. This phenomenon, not observed in previous experiments \cite{Wei2000, Lutz2004} is an additional feature of the supported manifold $S^{1}$, since the compactness of the circle implies the finite-size of the system, as well as a collective state where all particles behave as one ring-shaped particle with a rotational Brownian motion. After this stage, the geometrical regime appears as a consequence of the finite length of the circle.

Our results also allowed us to estimate expressions for the three transition times between the temporal regimes, and a novel expression for the single file mobility factor, $F$.  The first transition time turned out to be inversely proportional to the square value of the density, whereas the second transition time is proportional to the radius square of the circle. However, both transition times are inversely proportional to the value of the interaction potential divided by the thermal energy, $\beta u(r_{\rm max})$, where $r_{\rm max}$ corresponds to the position of the first peak of the angular distribution function. Besides that, we find that the third transition time is independent of the interaction potential and temperature, and becames proportional to the square of the number of particles in the colloidal system. Regarding the  analytical structure of the transitions times and the behavior of the MSAD, we deduce the single file mobility $F$. In this case, we found that $F$ is inversely proportional to the product between the particle density and the square root of $\beta u(r_{\rm max})$. In contrast to other expressions for the mobility coefficient (for instance, the one in Ref. \cite{Kollmann2003}), our expression has a direct connection with the interaction potential, which can also be used to probe directly the interaction between the particles. We also provided an accurate analytical expression that can easily reproduce either experiments or simulations in terms of the transition times.

Our approach can be extended in various directions. For instance, the Ermack-McCammon algorithm implemented on the circle can be  used  for other type of interaction potentials. In addition, the full formulation of the Ermack-McCammon algorithm on curved manifolds can be used to explore the behavior of the dynamics of a tracer  particle in different geometries, where curvature-induced inhomogeneities are relevant. Recently, there has been an important interest in the study of active particle systems, since they appear in a broad range of contexts, in particular, the single-file diffusion of active colloids in confined geometries \cite{Ebbens-2016, Zhang-2019, Dolai-2020} could be modeled with our approach including the active internal degree of freedom in our Ermack-McCammon algorithm on curved manifolds. Furthermore, by trapping paramagnetic particles along a closed curve using optical tweezers, controlling the temperature using a second laser beam \cite{Blicke}, and allowing changes from the external magnetic field as a thermodynamic variable, it might be possible to build and design colloidal heat engines with a few bodies \cite{Ignacio-2017}, where our approach could be essential. Finally, we could extend our findings by introducing the number density, $\rho=N/2\pi R$, allowing us to rewrite the transition times and the single-file mobility in a more general description.

\begin{acknowledgments}
A. V. B. acknowledges the financial support provided by Conacyt (CVU 417675). Further financial support by Conacyt (Grants Nos. 237425 and 287067, and Red Tem\'atica de la Materia Condensada Blanda) is gratefully acknowledged. The authors also thank to the General Coordination of Information and Communications Technologies (CGSTIC) at Cinvestav for providing HPC resources on the Hybrid Cluster Supercomputer ``Xiuhcoatl", which have contributed to the research results reported in this paper. R. C. P. also acknowledges the financial support provided by the Marcos Moshinsky fellowship 2013-2014. A. O. A. and P. T. acknowledge support from the European Research Council (ERC) under grant agreement no. 811234.
\end{acknowledgments}

\appendix

\section{First transition time $\tau_{d}$ from perturbation analysis}\label{app}
Here, we provide a perturbation analysis in order to determine an expression for the first transition time $\tau_{d}$. The key observation is that for a large value of the amplitude $\Gamma$, and a large value of the number of particles  $N$, the value for the first peak of the angular distribution function is approximately $\phi_{\max}\approx \frac{2\pi }{N}$. This feature, allows us to introduce the hypothesis that in this situation the particles in the system are localized around the mechanical equilibrium positions with angles  $\phi_{i}^{\rm eq}=\frac{2\pi}{N}(i-1)$ for $i=1, \cdots, N$. In addition, it is not difficult to show that the deterministic term $g_{i}:=\beta D_0 {\bf F}_{i} \cdot {\bf T}_{i}\tau$ of the  Ermack-McCammon algorithm (Eq. \ref{eq1}) is given by
\begin{eqnarray}
g_{i}\left(\phi_{ij}\right)=\sigma\left(\frac{\tau}{\tau_{\Gamma}}\right)\sum_{j\neq i}^{N}\frac{\cos\left(\frac{\phi_{ij}}{2}\right)\sin\left(\frac{\phi_{ij}}{2}\right)}{\left|\sin\left(\frac{\phi_{ij}}{2}\right)\right|^5},
\end{eqnarray}
where $\tau_{\Gamma}=16 R^4/(3\sigma^2 D_{0}\Gamma)$. It is clear that the function $g_{i}$ becomes zero at equilibrium positions since the forces between all the particles cancel out. Now, we carry out the perturbation $\phi_{ij}=\phi_{ij}^{\rm eq}+\eta_{ij}$, where $\phi_{ij}^{eq}=\frac{2\pi}{N}\left(i-j\right)$ and  $\eta_{ij}$ is a fluctuation around the equilibrium configuration. The expansion of $g_{i}$ in the first order of $\eta_{ij}$ is 
\begin{eqnarray}
g_{i}\left(\phi_{ij}\right)\approx -\left(\frac{\tau}{2R^{*}\tau_{\Gamma}}\right)\sum_{j\neq i}^{N}\mathcal{P}_{ij}\eta_{ij}, 
\end{eqnarray}
where the constants $\mathcal{P}_{ij}$ are given by
$\mathcal{P}_{ij}=\left(1+3\cos^2\left(\phi_{ij}^{\rm eq}/2\right)\right)/\left|\sin\left(\phi_{ij}^{\rm eq}/{2}\right)\right|^{5}.$ The fluctuation can be decomposed as $\eta_{ij}=\eta_{i}-\eta_{j}$, with $\eta_{i}=\delta{\bf r}(\tau)\cdot {\bf T}_{i}(\rm \phi_{i}^{\rm eq})$ satisfying the fluctuation-dissipation theorem  $\left<\eta_{i}\eta_{j}\right>=2\delta_{ij}D_{0}\tau$.

Next, we compute the first term of (\ref{eq22}), equivalent to $\langle g_{i}^2 \rangle$, using $\left<\eta_{ij}\eta_{ik}\right>=2D_{0}\tau\left(1+\delta_{jk}\right)$ for $i\neq j$ and $i\neq k$, then 
\begin{eqnarray}
\langle g_{i}^2 \rangle=2D_{0}\tau\left(\frac{\tau}{2R^{*}\tau_{\Gamma}}\right)^2\left[\left(\sum_{j\neq i}\mathcal{P}_{ij}\right)^2+\sum_{j\neq i}\mathcal{P}_{ij}^2\right].\nonumber\\
\end{eqnarray}
For simplicity let us take  $i=1$ for the tagged particle, and $j=2$ and $j=N$ for the first neighboring particles. Thus, for large $N$ one has $\left(\sum_{j}\mathcal{P}_{ij}\right)^2+\sum_{j}\mathcal{P}_{ij}^2\simeq 96 N^{10}/\pi^{10}$. Using this approximation, the mean-square angular displacement turns out to be
\begin{eqnarray}
\langle \left[ \Delta s_{i} (\tau) \right]^2 \rangle  \approx 2 D_{0} \tau\left[1+24\left(\frac{\tau N^{5}}{R^{*}\tau_{\Gamma} \pi^{5}}\right)^2 \right].\label{eq33}
\end{eqnarray}

Now, we choose for the interaction amplitude the value $\Gamma=\beta u(r_{\rm max})\left(r^{*}_{\rm max}\right)^{3}$, where approximately $r_{\rm max}^{*}=2\pi R^{*}/N$ according to the observation performed at the beginning of the appendix. The first transition time occurs  when the second term inside the square parenthesis in (\ref{eq33}) is around $1$. Therefore, one finds
\begin{eqnarray}
\frac{D_{0}\tau_{d}}{\sigma^2}=\frac{2\pi^2}{3\sqrt{24}}\left(\frac{R^{*}}{N}\right)^2\left(\beta u(r_{\rm max})\right)^{-1}.
\end{eqnarray}
The contribution of the next neighbors does not change the value $\mathcal{C}_{\rm th}=2\pi^2/(3\sqrt{24})\approx 1.3$  significantly.
Notwithstanding,  the factor $\mathcal{C}_{\rm th}$ obtained with the perturbation theory differs slightly from the simulation result, shown in Fig. (\ref{fig2}), the transition time has the correct structural dependence $\left(\frac{R^{*}}{N}\right)^2\left(\beta u(r_{\rm max})\right)^{-1}$.

\bibliography{SFDCircle}
\end{document}